\documentclass[a4paper,11pt]{article}

\usepackage{jcappub} 

\usepackage[T1]{fontenc} 
\usepackage{color,xcolor}

\newcommand\prl{{\it Phys. Rev. Lett.}}
\newcommand\prd{{\it Phys. Rev.}\rm\ D}
\newcommand\jcap{{\it JCAP}}
\newcommand\apjs{{\it ApJS}}
\newcommand\apj{{\it ApJ}}
\newcommand\aap{{\it A\&A}}
\newcommand\araa{{\it Annual Review of Astronomy and Astrophysics}}

\newcommand\gr{$\gamma$-ray}
\newcommand\fermi{{\it Fermi}}

\title{A possible blazar spectral irregularity case caused by photon--axionlike-particle oscillations}

\author[a]{Jianeng Zhou,}
\author[b,a,1]{Zhongxiang Wang,\note{Corresponding author.}}
\author[c]{Feng Huang,}
\author[a]{Liang Chen}

\affiliation[a]{Shanghai Astronomical Observatory,Chinese Academy of Sciences\\
80 Nandan Road, Shanghai 200030, China}
\affiliation[b]{Department of Astronomy, School of Physics and Astronomy, Key Laboratory of Astroparticle Physics of Yunnan Province, Yunnan University, Kunming 650091, China}
\affiliation[c]{Department of Astronomy, Xiamen University,
Zengcuo'an West Road, Xiamen 361005, China}

\emailAdd{zjn@shao.ac.cn}
\emailAdd{wangzx20@ynu.edu.cn}
\emailAdd{fenghuang@xmu.edu.cn}
\emailAdd{chenliang@shao.ac.cn}

\abstract{We report detection of a line-like feature in the $\gamma$-ray 
spectrum of the blazar B0516$-$621, for which the data obtained with the Large 
Area Telescope onboard {\it Fermi Gamma-Ray Space Telescope (Fermi)} are
analyzed. The feature is at $\sim$7\,GeV 
and different analyses are conducted to check its real presence. We 
determine that it has a significance of 2.5--3.0$\sigma$, and cautiously note
the presence of possible systematics in the data which could reduce 
the significance. This putative feature is too narrow to be
explained with radiation
processes generally considered for jet emission of blazars. Instead,
it could be a signal due to the oscillations between photons and axion-like 
particles (ALPs) in the source's jet. We investigate this possibility by
fitting the spectrum with the photon-ALP oscillation model, and find that
the parameter space of ALP mass $m_a\leq 10^{-8}$\,eV and the coupling
constant (between photons and ALPs) 
$g_{a\gamma}$=1.16--1.48$\times 10^{-10}$\,GeV$^{-1}$
can provide a fit to the line-like feature, while the magnetic field at the
emission site of $\gamma$-rays is fixed at 0.7\,G. The ranges for $m_a$ and 
$g_{a\gamma}$ are in tension with those previously obtained from several  
experiments or methods, but on the other hand in line with some of the others.
This spectral-feature case and its possible indication for ALP existence 
could be checked from similar studies of other blazar systems and also suggest
a direction of effort for building future high-energy facilities that would
have high sensitivities and spectral resolutions for searching for similar
features.
}

\begin{document}
\maketitle
\flushbottom

\section{Introduction}
\label{sec:intro}

It has been widely discussed that axion-like particles (ALPs), 
as a plausible
cold dark matter candidate, may reveal their existence with different 
signatures in astrophysical objects \cite{jr10,rin12}. These hypothetical
very light particles $a$ would interact with photons when pass
through external magnetic fields due to a two-photon coupling 
$a\gamma\gamma$ \cite{sik83,rs88}. For example, pulsars, the fast-rotating 
neutron stars with strong surface magnetic fields, might show
a differential delay in their pulsed emission signals from 
the magnetospheres \cite{mn93} 
or $\gamma$-ray spectral modulation because of high-energy photons going 
through the Galactic magnetic 
fields \cite{mch18}. In a core-collapse supernova (SN), significant amount
of ALPs could be produced, and they would be converted to 
high-energy $\gamma$-ray photons in the Galactic magnetic fields \cite{sn1987a}.
This scenario led to the search for $\gamma$-ray emission from SN1987A, 
a nearby SN event in the Large Magellanic Cloud. Blazars, as subclass
sources of Active Galactic Nuclei (AGN), contain a jet pointing 
close to our line of sight. The relativistic beaming effect makes the jets'
emission bright and highly variable. Because the magnetic field in a blazar 
jet can be of the order of
$\sim$1\,G, its emitted high-energy photons could be
converted to ALPs and thus a non-thermal smooth spectrum would be 
distorted  \cite{hs07,hs207}. In addition, very-high energy (VHE) $\gamma$-ray 
photons from distant blazars are severely absorbed because of the interaction 
with the extragalactic background light (EBL). However, as they pass through
magnetic fields of different strengths at galactic, intra-galactic, 
and intra-cluster regions, the conversion between the photons and ALPs could
reduce the absorption effect as ALPs would travel freely,
increasing the detectability of distant blazars at VHE 
bands \cite{drm07,dgr11,hmm+12}.

Considering an external magnetic field with component $B_T$ transverse to 
the propagation direction of photons, it can be shown that there is a 
critical energy $E_c$ for photons of energy $E$, 
\begin{equation}
E_c = \frac{|m_a^2-\omega_{\rm pl}^2|}{2 g_{a\gamma} B_T}\sim 0.25\ {\rm GeV}\ \frac{|m_a^2-\omega_{\rm pl}^2|}{(10^{-8}\ {\rm eV})^2} (\frac{10^{-3}\ {\rm G}}{B_T}) (\frac{10^{-11}\ {\rm GeV}^{-1}}{g_{a\gamma}}),
\end{equation}
where $m_a$ is the ALP mass, $\omega_{\rm pl}$ is the plasma frequency
(determined by the electron density $n_e$, $\omega_{\rm pl}\sim 0.037\sqrt{n_e\, {\rm cm}^{-3}}$), and $g_{a\gamma}$ is the coupling constant between photons
and ALPs (e.g., ref. \cite{hs07,dmr08,hmm+12}). Around $E_c$, significant 
photon-ALP mixing is expected. Specifically for
blazars, high-energy photons are emitted from a blob region $\sim$0.01--0.1\,pc
away from the central supermassive black hole and the photons travel through
a jet with a length of $\sim$kpc and nearly pointing to the Earth. The magnetic 
field and electron density of the jet decrease as a function of the distance $r$
from the blob along the jet (often assumed to be $\sim r^{-1}$ for the former 
and $\sim r^{-2}$ for the latter). Adding photon-ALP oscillations within 
this jet,
the observed high-energy spectrum would appear distorted (e.g., \cite{trg15}).
Previously, pieces of possible evidence for such photon-ALP oscillations
were seen in spectra of the jets at GeV energies 
(e.g., \cite{mr13,aje+16,mal+18})

Since the launch of {\it Fermi Gamma-ray Space Telescope (Fermi)} in 2008, 
the Large Area Telescope (LAT) onboard it has been scanning the whole sky
for detecting \gr\ sources at GeV energies and collecting data for different
types of detected sources. From 10-yr data collected with LAT, more than 3000
blazars (the dominant sources in the sky) have been found. The details
of the sources were reported in \fermi\ LAT Fourth Source Catalog (4FGL) 
\cite{4fgl20} and Data Release~2 (4FGL-DR2) \cite{4fgldr2}. A rich amount
of spectral data are available for searching for possible photon-ALP oscillation
signals. In this paper, we report a line-like feature found in the 
spectrum of a blazar, QSO~B0516$-$621 (e.g., ref. \cite{sha+12}). 
Standard radiation processes considered 
for jets' emission cannot explain this feature. After our thorough searches for
possible explanations, we realized that it can be 
considered as a photon-ALP oscillation signal. Detailed studies of
this case was conducted.  In the following section~\ref{sec:ana}, we describe
the analysis of the \fermi\ LAT data to establish detection of the line-like 
feature.
We introduce the photon-ALP oscillation model we used to fit the spectrum
of B0516$-$621 and present the fitting results in section~\ref{sec:alp}.
In section~\ref{sec:dis}, the results are discussed.

\begin{figure}[tbp]
\centering 
\includegraphics[width=.45\textwidth]{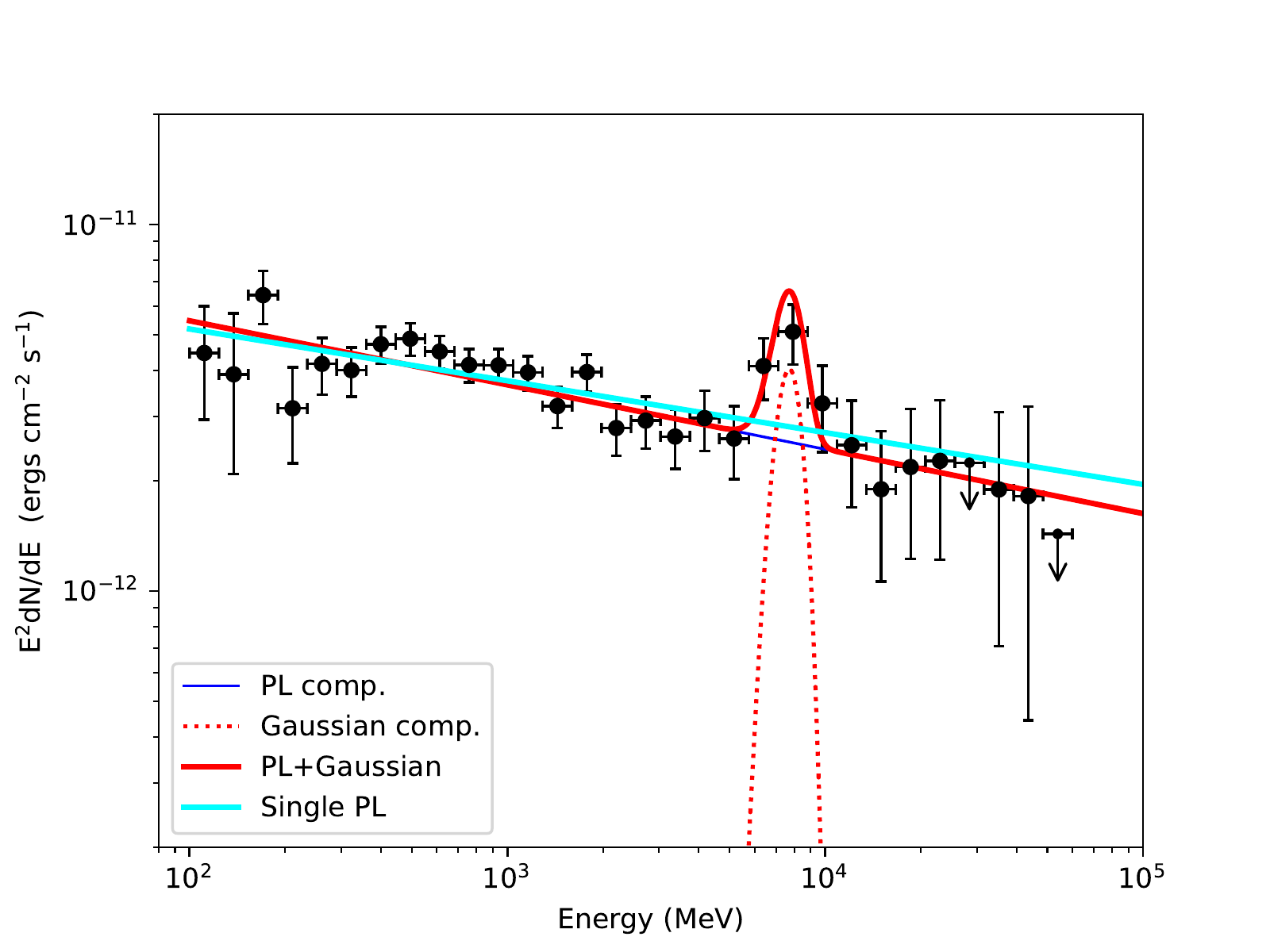}
\hfill
\includegraphics[width=.45\textwidth]{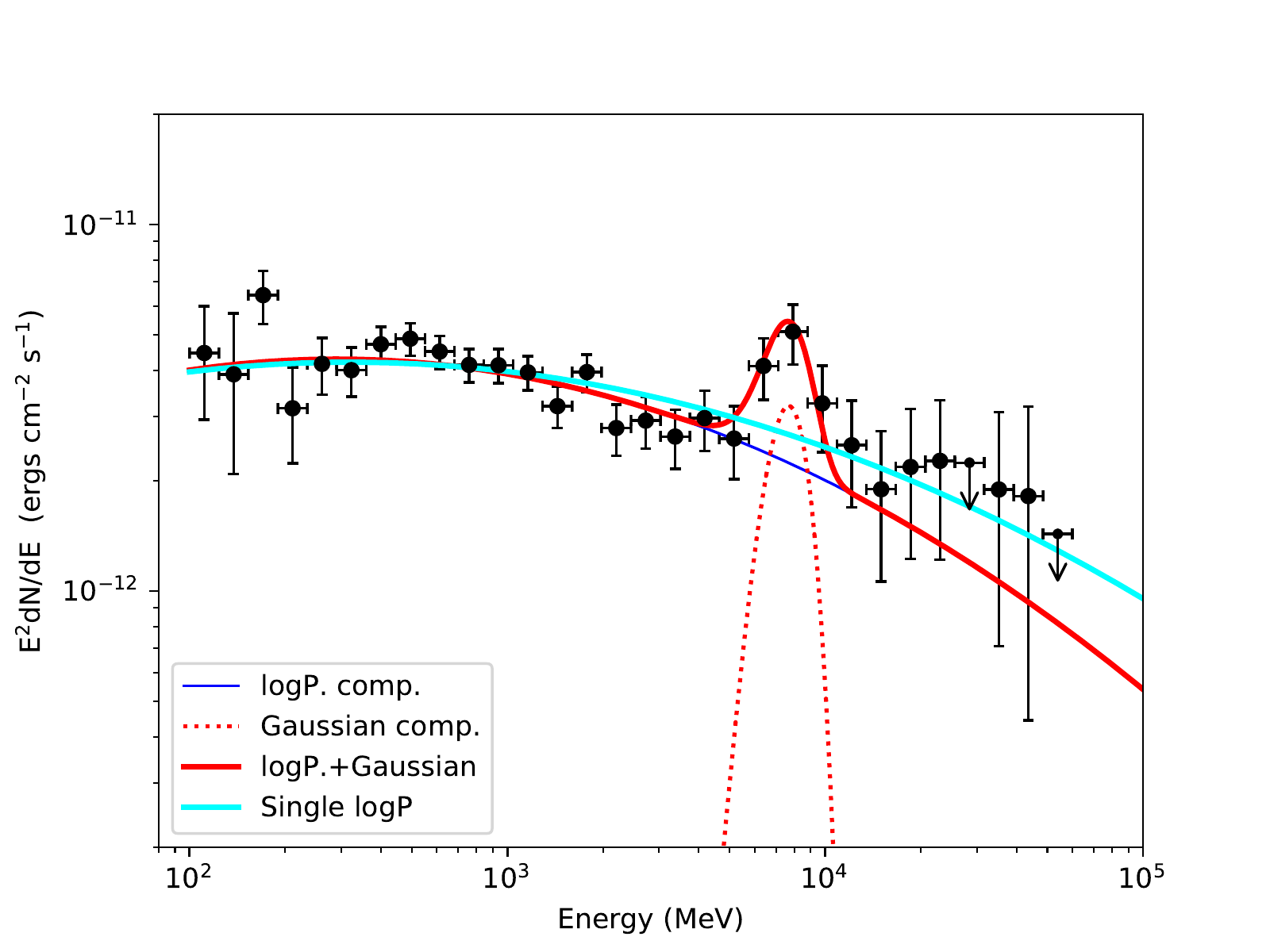}
\caption{$\gamma$-ray spectrum of B0516$-$621 in energy range of 0.1--60\,GeV,
obtained with \fermi\ LAT.  A line-like feature at $\sim$7~GeV is seen. 
The spectrum is fit with the model
of a power law plus a Gauss ({\it left} panel) and that of a LogParabola plus
a Gauss ({\it right} panel). 
\label{fig:spec}}
\end{figure}

\section{$\gamma$-ray data analysis for B0516$-$621}
\label{sec:ana}

\subsection{\fermi\ LAT data and source model}
LAT has been scanning the whole sky at energies from 20\,MeV to $>$300\,GeV. We
used the data in the energy range of 0.1--300\,GeV from 2008 August 4 to 2018 
August 4 (10 yrs) in our analysis. The events of Pass 8 `SOURCE' class 
(evclass=128, evtype=3) within a $20^{\circ} \times 20^{\circ}$ region 
of interest (RoI) were selected, which is centered at the target B0516$-$621 
(the \fermi\ LAT counterpart is named as 4FGL~J0516.7$-$6207). To avoid 
contamination from the Earth's 
limb contamination, we excluded the events with zenith angles $> 90^{\circ}$. 
The latest instrument response function (IRF) 
{\tt P8R3\_SOURCE\_V3}
and analysis tool 
Fermitools 2.0.8 were used.

\begin{figure}[tbp]
\centering 
\includegraphics[width=.45\textwidth]{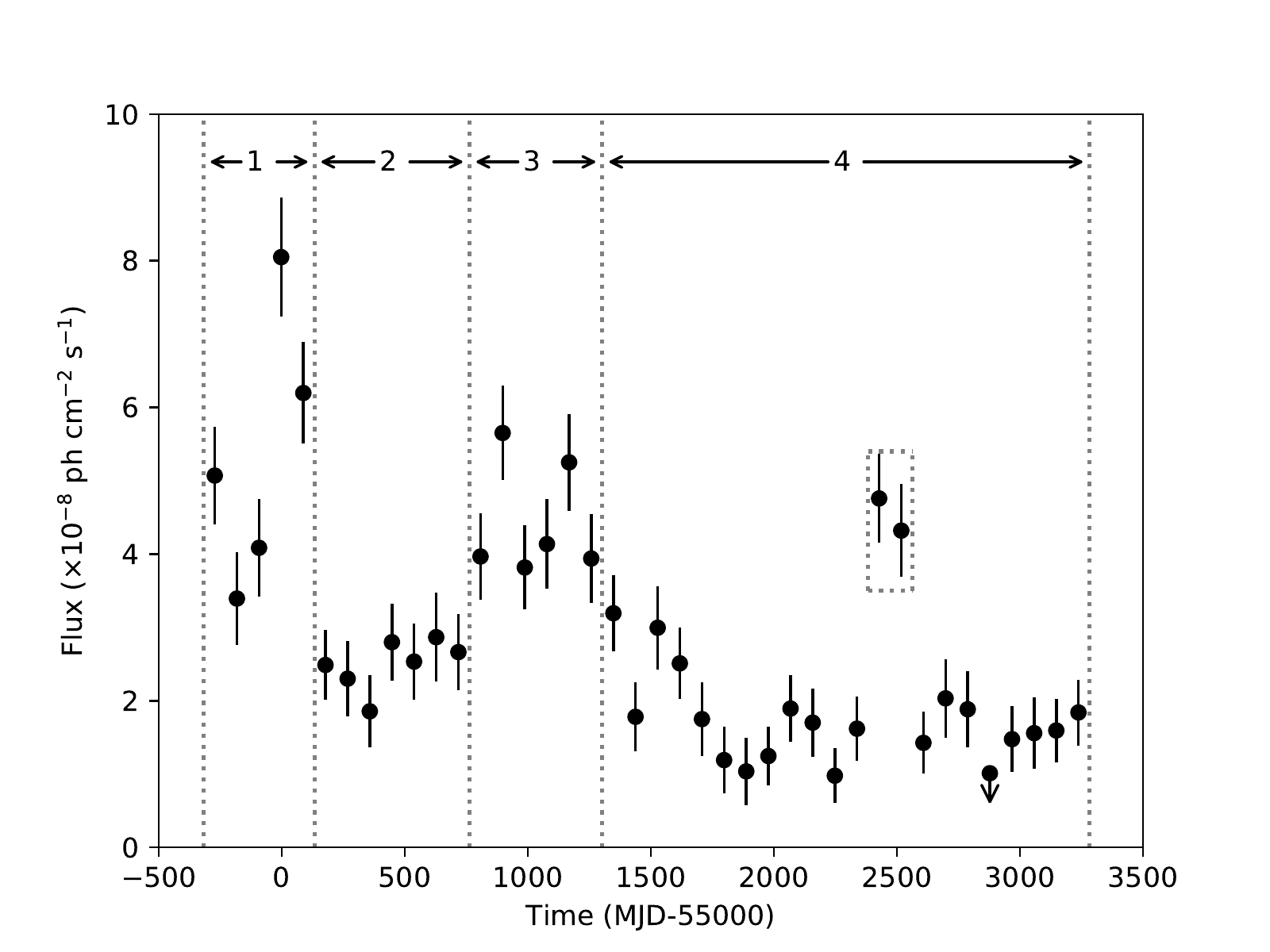}
\hfill
\includegraphics[width=.45\textwidth]{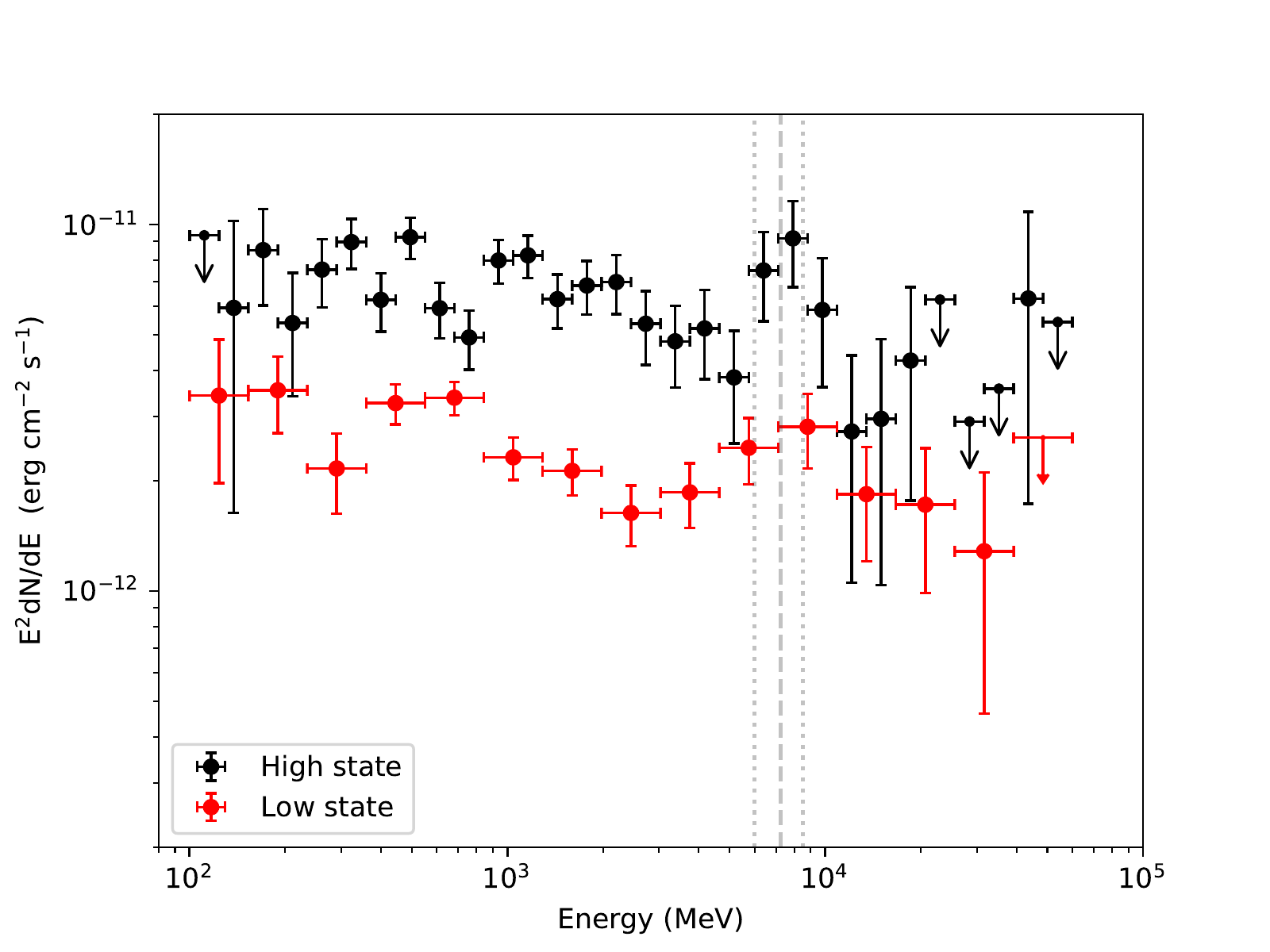}
\caption{{\it Left} panel: 90-day binned light curve of B0516$-$621. Based on
the light curve, four time periods are defined: Time periods 1 \& 3 plus two
data points (around MJD~57500 in the dotted box) are
in the high state, while the remaining time periods (mainly 2 \& 4) are
in the low state. {\it Right} panel: Spectra obtained from the data in 
the high state (black data points) and low state (red data points). The 
position of the line-like feature is marked by vertical dashed and dotted
lines.
\label{fig:lc}}
\end{figure}

We constructed a source model using the recently released 
4FGL-DR2 \cite{4fgldr2}. Sources listed in 4FGL-DR2 within the RoI were
included in the source model, 
whose spectral models given in the catalog were used.
Spectral parameters of the sources were basically fixed at the catalog values,
but for the variable sources or those within 5~degrees away from our target,
their normalizations were set free in our analysis.
The background Galactic and extragalactic diffuse 
spectral models (gll\_iem\_v07.fits and 
iso\_P8R3\_SOURCE\_V3\_v1.txt
respectively) were also included in the source model. Their
normalizations were set as free parameters.

In the binned likelihood analysis presented in the following
sections, energy dispersion was enabled, although without which the spectral
distortion is limited to less than 5\%  in the 1--100\,GeV energy range\footnote{https://fermi.gsfc.nasa.gov/ssc/data/analysis/documentation/Pass8\_edisp\_usage.html}.
The controlling parameter {\tt edisp\_bins=$-$2} was set for the implementation
of the energy dispersion correction. (For comparison, the analysis without
energy dispersion enabled gave a slightly higher significance 
($= 3\sigma$) for the line-like feature we found.)

\subsection{Spectral analysis}
\label{sec:sa}

We obtained the spectrum of B0516$-$621 by performing the maximum likelihood 
analysis of the LAT data in evenly divided energy bins in logarithm. There
were few photons above 60~GeV, and we found 30 bins
between 0.1--60~GeV allow to show the spectrum relatively well without
having many flux upper limits at the high-energy end.
The obtained spectrum is
shown in Figure~\ref{fig:spec}, and a line-like feature is seen
near $\sim 7$\,GeV. 
\begin{figure}[tbp]
\centering 
	\includegraphics[width=.45\textwidth]{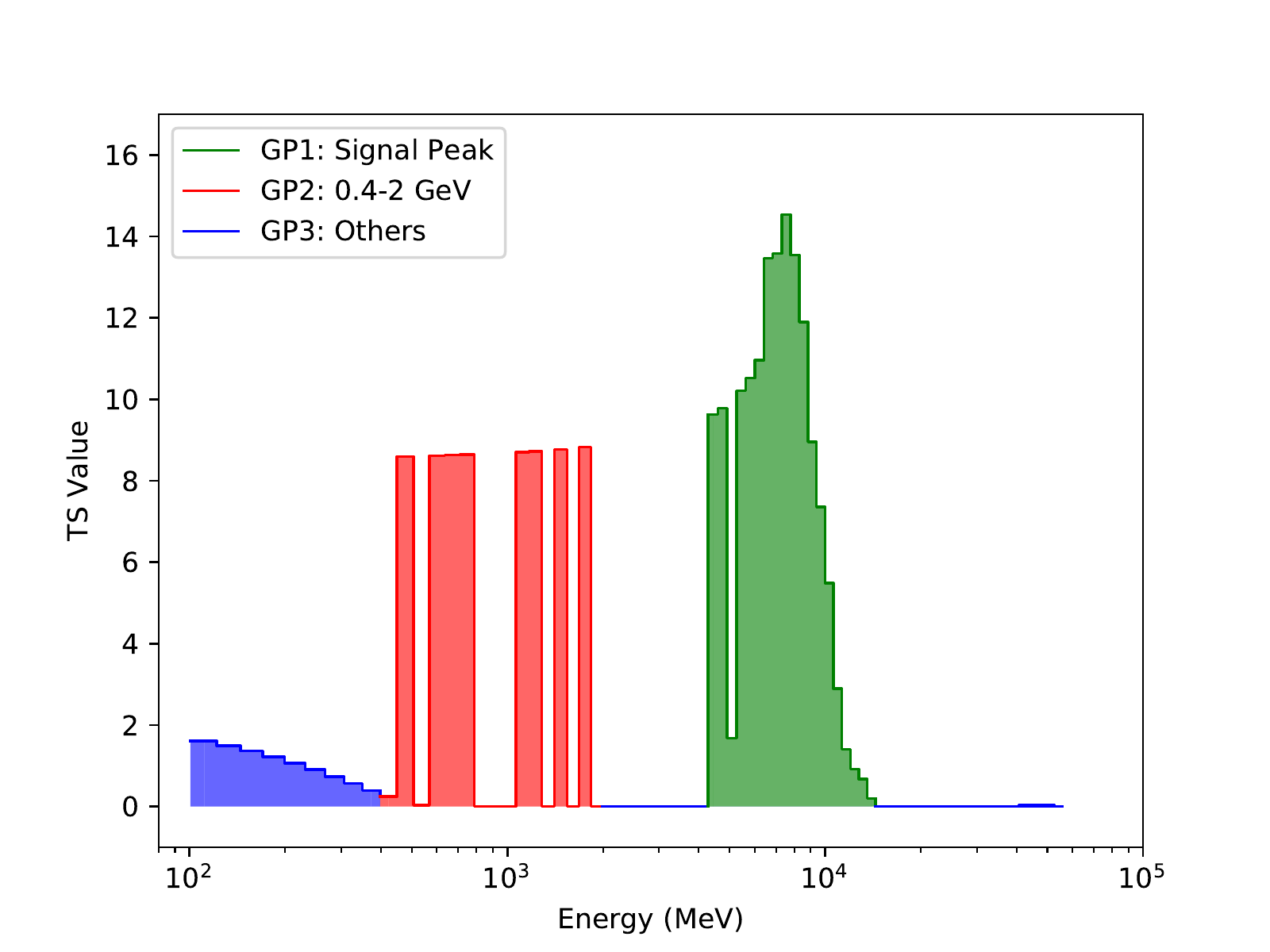}
	\includegraphics[width=.45\textwidth]{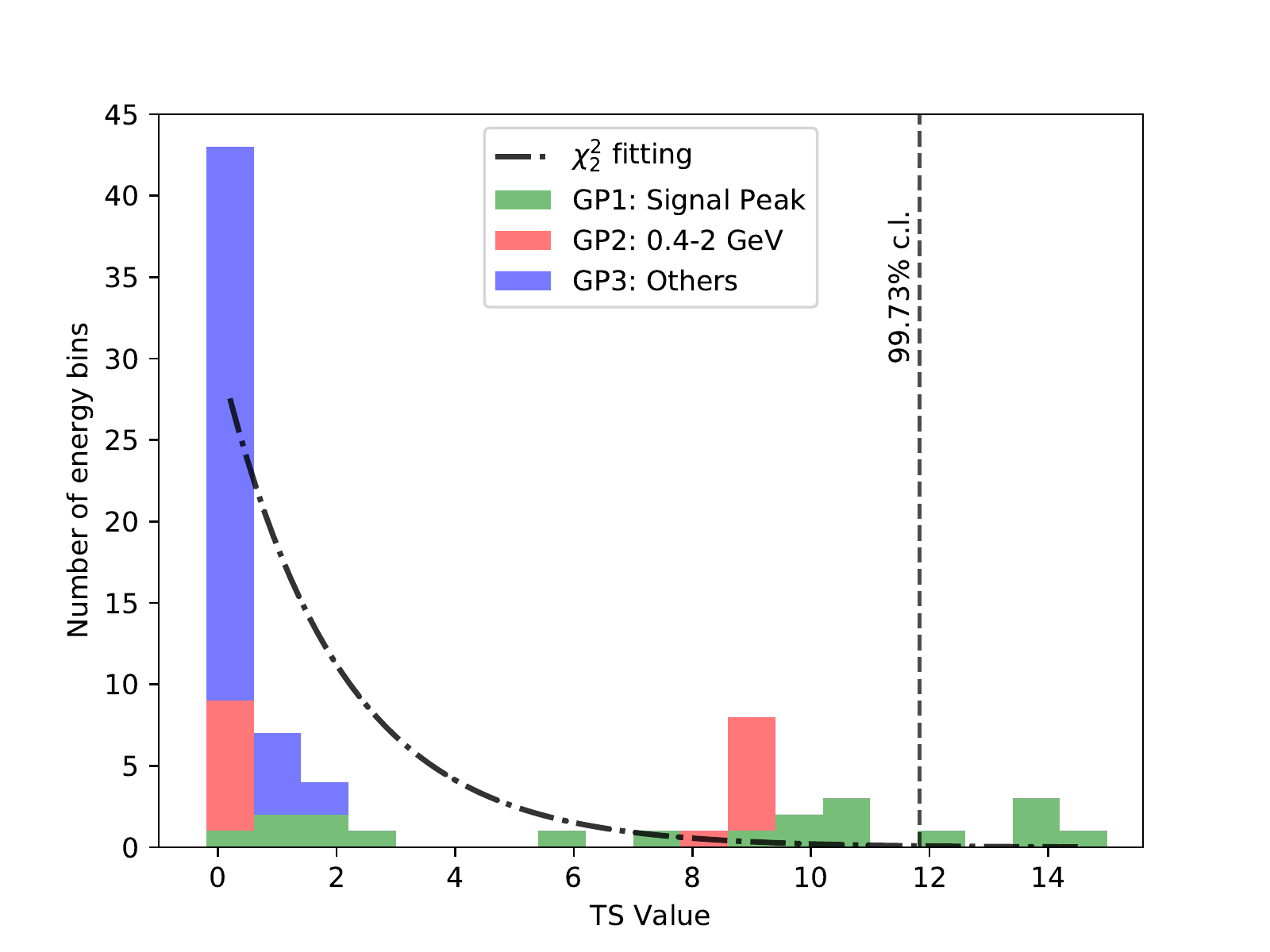}
	\caption{{\it Left:}
TS values for having an additional Gaussian component moved over the energy 
range of 0.1--60\,GeV, where the step for moving the Gaussian component 
	was set as the energy resolution at a considered energy.
	The maximum TS value is 14.54 at $\sim 7.5$\,GeV. Three groups of
	the data points---the signal peak around 7\,GeV, 0.4--2\,GeV, and
	the others having TS$< 2$---are marked with different colors. {\it Right:} distribution
	of the TS values shown in the left panel. A $\chi^2$ distribution (with
	2 degrees of freedom) is forced to fit the TS ditribution, and the 
	3$\sigma$-significance position from the $\chi^2$ distribution is marked by the dashed line. A few
	TS$\simeq 8$--10 results are notable, which likely represent the 
	systematic uncertainties in the analyzed data. 
\label{fig:sTS}}
\end{figure}

By assuming a power law $dN/dE \propto E^{-\Gamma}$ for the source, where
$E$ is photon energy and $\Gamma$ is photon spectral index, the likelihood 
analysis to the whole data in 0.1--60~GeV resulted in 
$\Gamma=2.14\pm0.02$.
This power law
does not fit the part around $\sim$7\,GeV in the spectrum.
In order to check if the addition of a narrow feature can improve the fit,
we added a Gaussian component to the power law,
$dN/dE \propto {\rm exp}[-(E-E_p)^2/(2\delta E^2)]$, where $E_p$ and
$\delta E$ are the peak energy and width of the component.
The improvement of having the Gaussian component can be estimated from
the test statistic (TS)
${\rm TS}=-2\log(L_{\rm pl}/L_{\rm plg})$, where $L_{\rm pl}$ and $L_{\rm plg}$
are the maximum likelihood values modeled with a power law and a power law 
plus a Gauss, respectively (e.g., ref. \cite{2fgl}). We found 
TS$=10.87$, which corresponds to 
2.50$\sigma$ significance for including the Gaussian component, 
as the TS value is considered to be distributed as $\chi^2$ with 3 degrees 
of freedom. In this fitting, we obtained 
$\Gamma=2.17\pm0.03$
and $E_p=7.54\pm0.84$~GeV
(see the left panel of Figure~\ref{fig:spec}).

Since more data have been collected with LAT for different sources,
it has been found that blazars' emission may be more generally described with a 
LogParabola function \cite{che14,4lac}, $dN/dE\propto \exp[-\alpha-\beta\log(E)]$, where
$\alpha$ is the spectral slope and $\beta$ describes the curvature of a 
spectrum. Using this function, we repeated the above analysis but with a
LogParabola function. We obtained 
$\alpha = 2.09\pm0.03$ and 
$\beta = 0.05\pm0.02$. 
Then adding a Gaussian function, we found 
TS$=13.94$,
which corresponds to 
2.97$\sigma$ significance, in which 
$\alpha = 2.13\pm0.01$,
$\beta = 0.06\pm0.01$, $E_p=7.25\pm0.18$~GeV, $\delta E=1.27\pm0.22$\,GeV
(see the right panel of Figure~\ref{fig:spec}). 

We conducted other analyses to check if this line-like feature 
would be due to uncertainties in the IRF or have an instrumental origin.
We obtained the source spectrum using the `CLEAN' event class (evclass=256),
which has a 1.3--2 times lower background rate above 3\,GeV than 
the SOURCE event class.
The spectrum is nearly the same as that obtained above, with no discernible 
differences. We also obtained the spectra of the $\gamma$-ray bright and stable 
pulsars, the Crab, Vela, and Geminga, from the same time period of the data and
with the same analysis settings as those in the above analysis for B0516$-$621. 
Their spectra appear as 
smooth curves with no similar line-like features seen.
From these analyses, we conclude that we have found a 
line-like feature at $\sim 7$\,GeV in the spectrum of B0516$-$621 and 
this feature is significant
at a $\sim 3\sigma$ level (also see Section~\ref{sec:swa}).

\subsection{Temporal analysis}
Blazars are highly variable sources, and B0516$-$621 exhibited strong
flux variations at $\gamma$-rays as well. In the left panel of 
Figure~\ref{fig:lc}, a 90-day binned light curve is shown, for which
we obtained the fluxes by performing the binned likelihood analysis to the
LAT data in each of 90-day time bins. In this analysis, the catalog sources 
within 5 degrees from the target were set to have free prefactors. Based on the 
light curve, we approximately divided it into four time periods. 
We considered that time periods 1 \& 3, and plus two data points 
around MJD~57500 (marked in a dotted box in Figure~\ref{fig:lc}), are
in a high state, and the remaining time periods (mainly 2 \& 4) are in
a low state. 
\begin{figure}[tbp]
\centering 
\includegraphics[width=.65\textwidth]{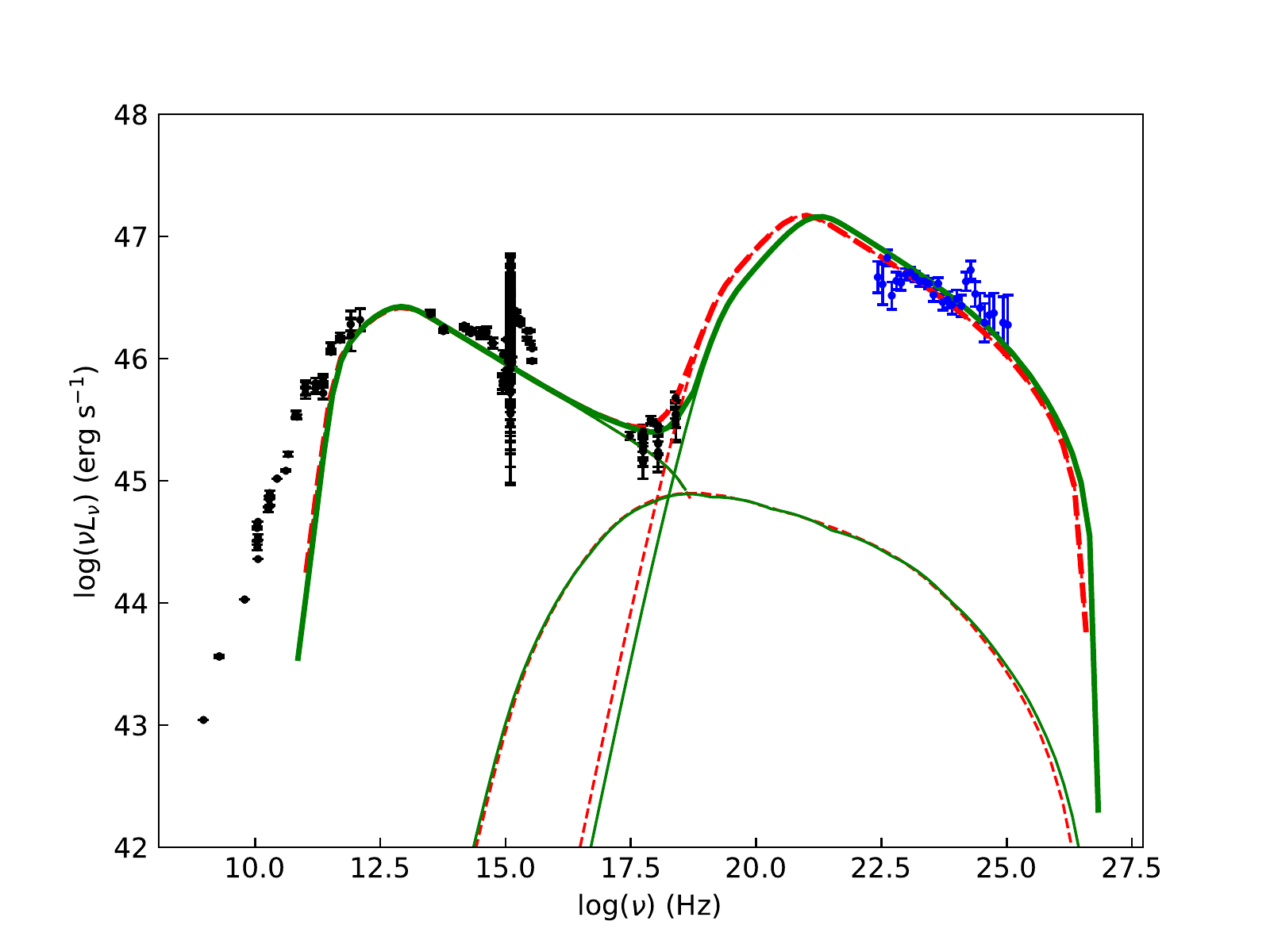}
\caption{
Broadband SED of B0516$-$621. The green curves and red curves are 
the model spectra with $B_T^0=0.7$\,G and $B_T^0=1$\,G, respectively.
The standard one-zone synchrotron plus Comptonization model is applied, 
while the left, middle, and right components are the synchrotron,
self-synchrotron Compton scattering, and external Compton scattering
emission, respectively.
\label{fig:bspec}}
\end{figure}

We then obtained spectra from the data in each of the two states, which
was for the purpose of checking if the line-like feature was only related to
any brightening activity of the source \cite{m501} or due to addition of
different spectral forms in different states. Note that because the source
was not sufficiently bright, we could only obtain decent spectra from
such two sets of the data. The spectra are shown in
the right panel of Figure~\ref{fig:lc}. For the spectrum in the low state,
we only divided the energy range into 15 bins, to maximize the number
of spectral data points (i.e., avoid to have too many flux upper limits).
The two spectra show main differences in the 1--4~GeV energy range, but
at the position of the line-like feature, both have upward data points deviating
from a single power law or LogParabola.
Given these and the relatively large flux uncertainties, no evidence is
seen for the non-presence of the line-like feature in the low state.
We tentatively conclude that the feature is present in emission of the source
all the time and not due to the addition of different spectral shapes in
different states.
\begin{figure}[tbp]
\centering 
\includegraphics[width=.36\textwidth]{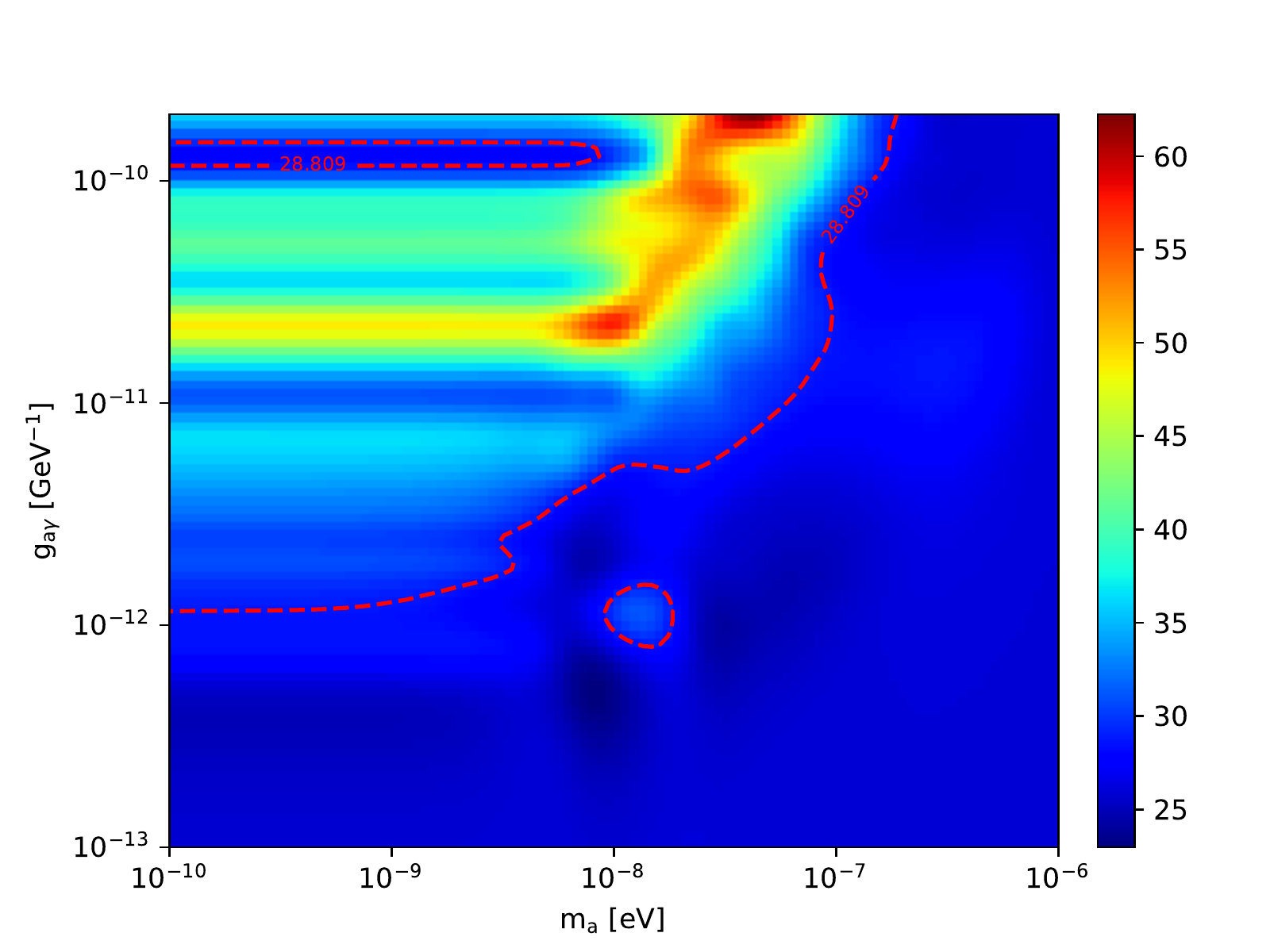}
\includegraphics[width=.32\textwidth,trim=0 -4cm 0 0]{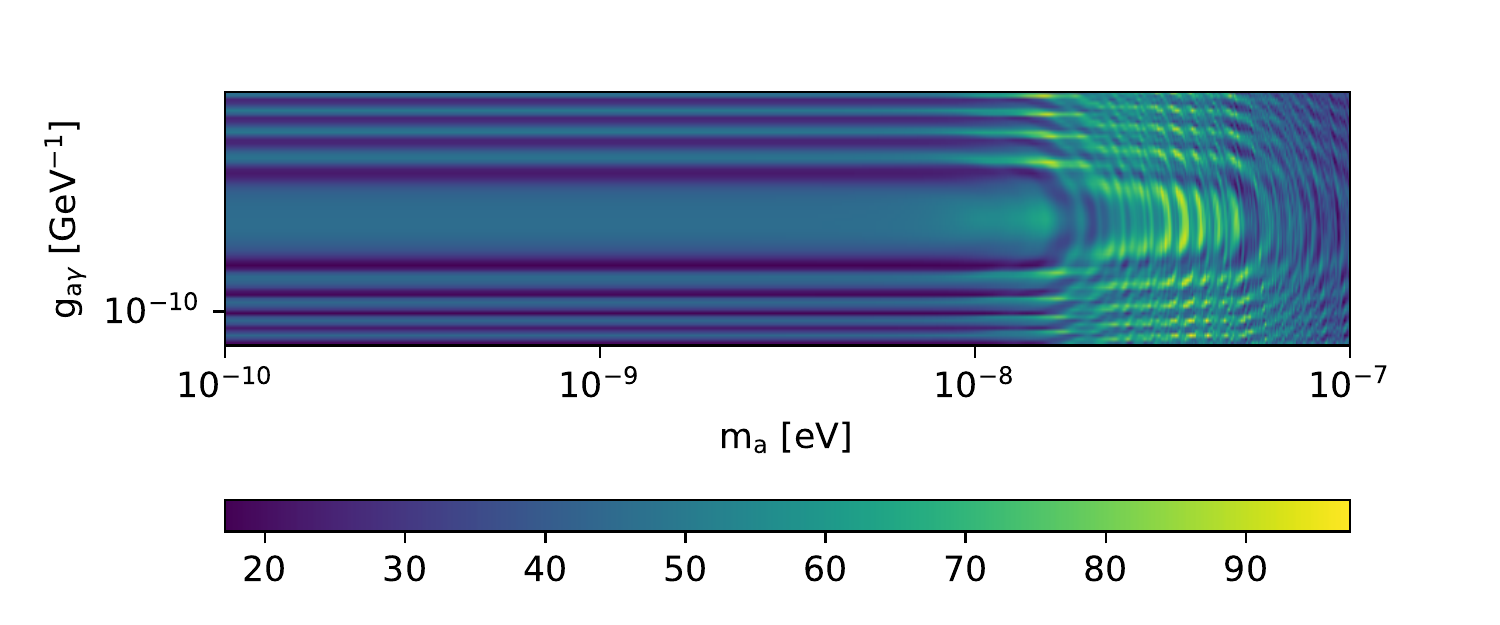}
\includegraphics[width=.28\textwidth]{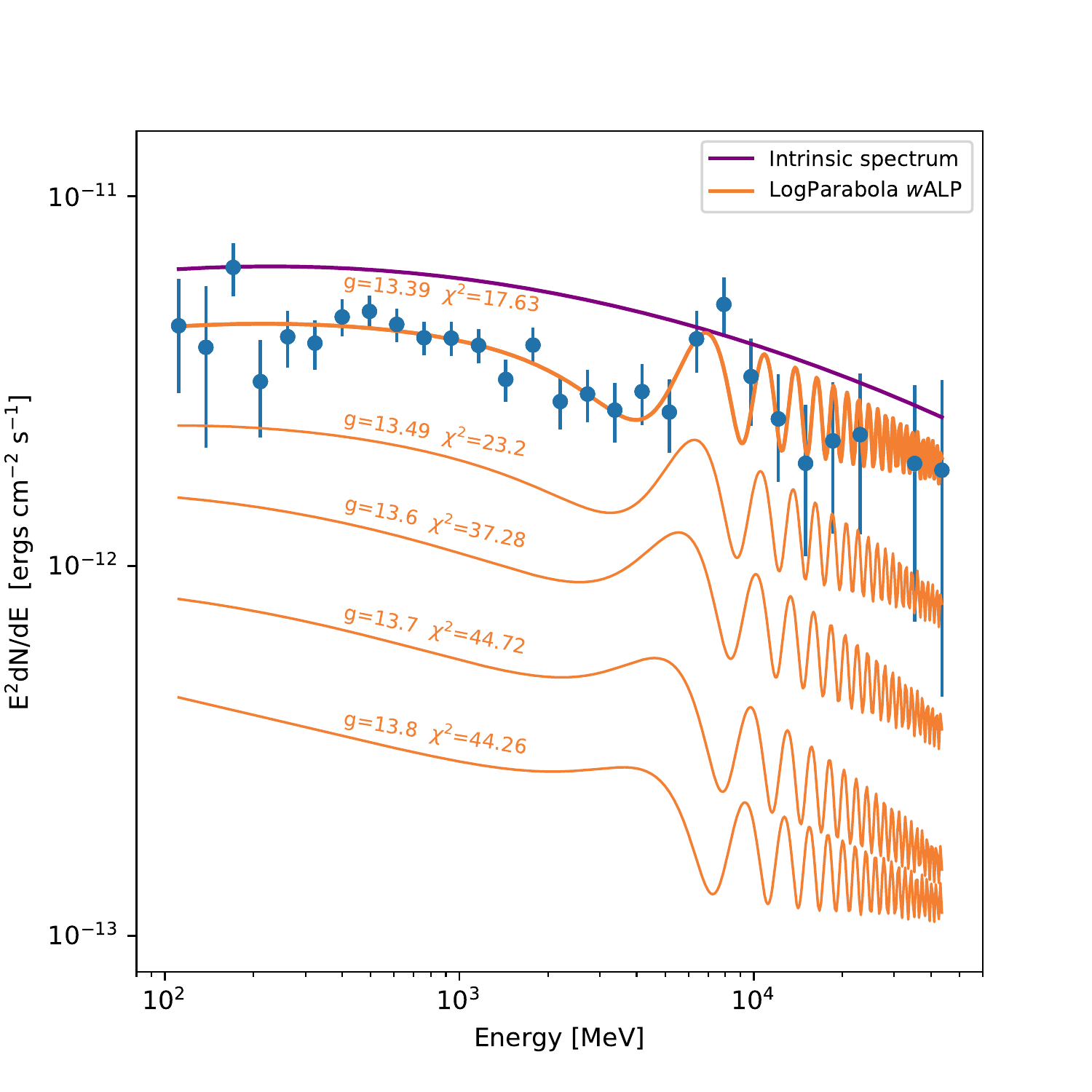}
\caption{{\it Left:} $\chi^2$ results in the space of $g_{a\gamma}$ versus 
$m_a$, for which $B_T^0=0.7$\,G.  Red dashed curves mark 
the 2$\sigma$ regions for the two parameters from our fitting, 
and the color bar indicates the $\chi^2$ value ranges. 
{\it Middle:} $\chi^2$ results for the region of 
$10^{-10}\leq m_a\leq 10^{-8}$\,eV and 
	$1.16\times 10^{-10}\leq g_{a\gamma}\leq 1.48\times 10^{-10}$\,GeV 
with fine grids (the upper left
corner of the left panel, indicated by the red-dashed curve). The color bar
indicates the $\chi^2$ value ranges. {\it Right:} A set of example 
model fits with $m_a=10^{-9}$\,eV (positions along the Y-axis 
are adjusted for clarity) to the observed $\gamma$-ray spectrum.
The $g_{a\gamma}$ value and resulting $\chi^2$ value of each fit
are given. The intrinsic spectrum for the uppermost model fit 
(the best one) is shown as the purple curve.
\label{fig:fit}}
\end{figure}

\subsection{Spectral feature analysis}
\label{sec:swa}
As the spectrum of B0516$-$621 (Figure~\ref{fig:spec}) generally appears to 
deviate from a smooth curve, for example a possible weak feature also seen 
at $\sim 150$\,MeV, we further conducted analysis for searching for
other possible line-like features. Similar to the above analysis in
section~\ref{sec:sa}, we used a LogParabola to represent the continuum
component and a Gauss to represent a possible line-like feature.
The peak energy $E_p$ of the Gauss was moved over the energy range
of 0.1--60\,GeV, with the moving step being the energy resolution at the energy
considered (similar to the sliding energy analysis; e.g., ref. \cite{br2011}).
The energy resolution was set as the 68\% containment half width of 
the reconstructed incoming photon energy. The TS value for having an
additional Gaussian component at a given energy compared with a single 
LogParabola was calculated 
(cf., section~\ref{sec:sa}). 

The obtained TS values are shown in the left panel of
Figure~\ref{fig:sTS}. There are a few TS$\simeq 8$--9 data points 
between 0.4--2\,GeV, suggesting low-significant ($\sim 2\sigma$; estimated
from a $\chi^2$ distribution with 2 degrees of freedom) Gaussian features.
Then a group of slightly higher TS values are present around 7\,GeV, with
the maximum value of $\sim$ 14.54 at $\sim$ 7.5\,GeV 
(corresponding to a significance of $\sim 3.2\sigma$). This group of
the data points are presumbly caused by the line-like feature.
We show the distribution of the TS values in the right panel of
Figure~\ref{fig:sTS}, and note that a $\chi^2$ distribution cannot well 
fit the TS distribution. The possible reason for the poor fit is 
that nearly
half of the data points in 0.4--2\,GeV do not have low TS values
(e.g., $\sim$0--6), in addition to the group of the data points around the
line-like feature. The deviation between the two distributions
likely indicates the presence of systematic uncertainties in 
the analyzed data, which would thus reduce the significance of
the observed feature.

In any case, this analysis indicates that there are no other similarly 
significant
features in emission of the source, and points to the same energy location 
where the line-like feature was identified.  

\section{Photon-ALP oscillation model and spectral fitting}
\label{sec:alp}
As discussed below in the following section~\ref{sec:dis}, it is 
hardly to have a mechanism in the generally considered frameworks of the blazar 
radiation processes (for a recent review focusing on their \gr\ emission, see,
e.g., ref. \cite{ms16}). Instead, the line-like feature is
a possible piece of evidence that indicates photon-ALP oscillations in
jets of AGN. We thus conducted detailed study for fitting the spectrum
of B0516$-$621 with photon-ALP oscillations included.

We used the photon-ALP oscillation model in blazar jets described
in the ref. \cite{mmc14}\footnote{https://gammaalps.readthedocs.io/en/latest/}. Along a jet launched from a supermassive
black hole in the center of a galaxy, its transverse magnetic field 
$B_T$ and electron density $n_e$ decrease as functions of distance 
$r$ (from the black hole) following $B_T(r)=B_T^0(r/r_0)^{\alpha_j}$,
$n_e(r)=n_e^0(r/r_0)^{\beta_j}$, where $r_0$ is the distance of the emission 
site of $\gamma$-rays from the black hole. Following ref. \cite{mmc14}, we set
$\alpha_j=-1$ and $\beta_j=-2$.  

By fitting the broadband spectrum energy distribution (SED) 
of B0516$-$621 \cite{abd+10} with 
the typical blazar radiation processes \cite{che18} (i.e., synchrotron for 
radio--optical/X-ray emission and inverse Compton scattering for 
X-ray--$\gamma$-ray emission), the parameters of 
$B_T^0=0.7$\,G, 
the Doppler beaming factor $\delta=20.3$, and the size of the emission region
$R\simeq 6.7\times 10^{16}$\,cm (radius of a homogeneous sphere) were required 
(see Figure~\ref{fig:bspec};
the redshift $z=1.3$; \cite{sha+12}). 
We assumed the bulk Lorentz factor
$\Gamma_b\simeq\delta$ and the viewing angle 
$\theta\simeq 1/\Gamma_b\simeq 2.8^{\circ}$. The distance of the emission
site was estimated to 
be $r_0\simeq R/(1/\Gamma_b)\simeq 1.3\times10^{18}$\,cm (or $\simeq$0.4\,pc;
e.g., ref. \cite{cz21}).
For the electron density at $r_0$, we assumed $n_0=5\times 10^4$\,cm$^{-3}$
(see, e.g., ref.~\cite{trg15}). 
The jet ended at 1~kpc from the black hole.
\begin{figure}[tbp]
\centering 
\includegraphics[width=.45\textwidth]{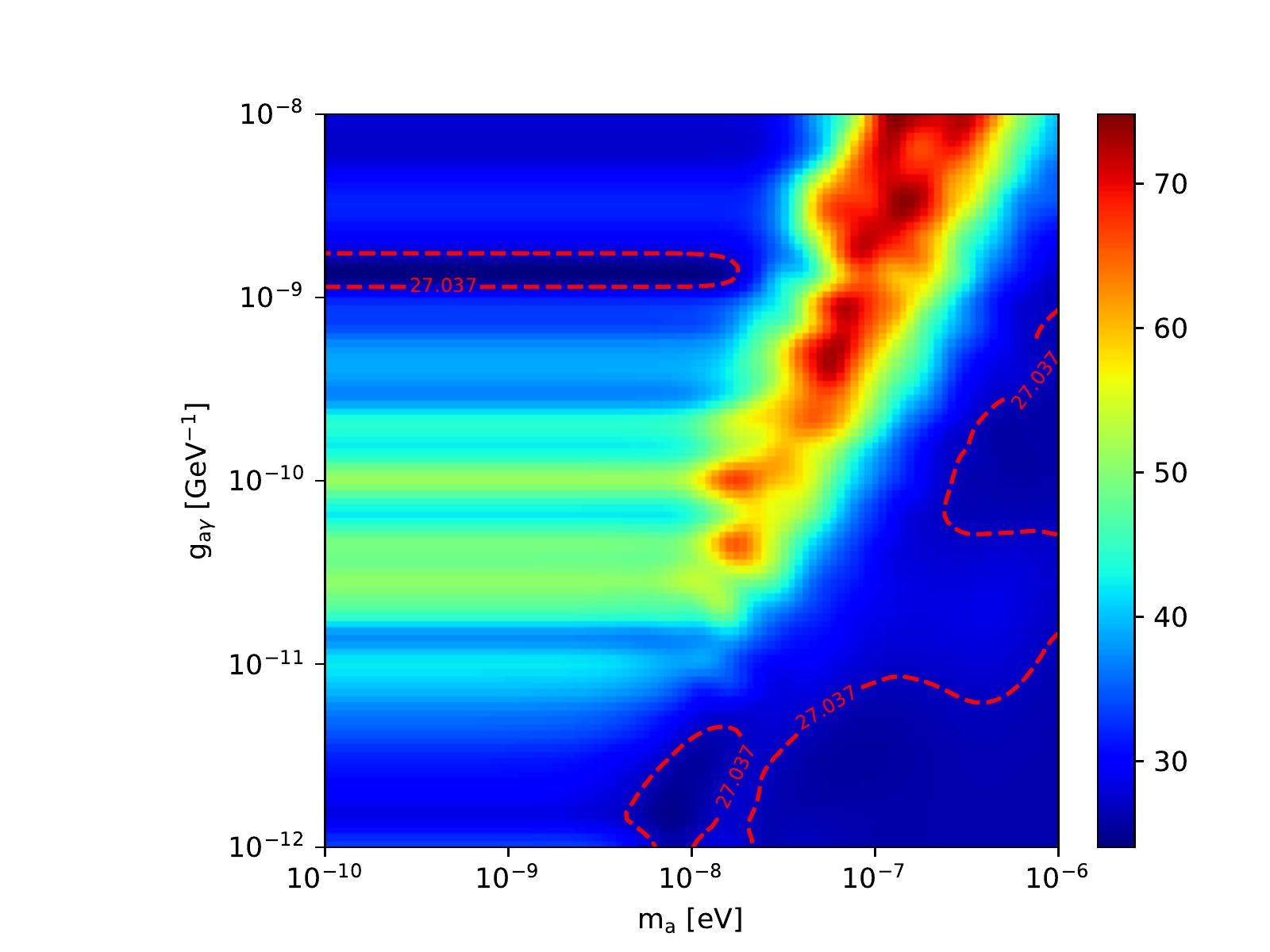}
\caption{$\chi^2$ results in the space of $g_{a\gamma}$ versus 
$m_a$, for which $B_T^0=1$\,G.  Red dashed curves mark 
the 2$\sigma$ regions for the two parameters from our fitting, 
and the color bar indicates the $\chi^2$ value ranges. 
\label{fig:b1g}}
\end{figure}

To fit the observed $\gamma$-ray spectrum, we input spectra
of a LogParabola form, the presumed intrinsic emission at
the $\gamma$-ray emission
site. The emission went through the jet and was oscillated due to 
the photon-ALP coupling. Each output spectrum 
was compared to the observed
$\gamma$-ray spectrum. A $\chi^2$ test was used to find the best-fit output
spectrum. In this fitting process, the magnetic field was fixed, while
the spectral parameters ($\alpha$, $\beta$, and prefactor) of 
the input spectra were set to have a large range
(i.e., treated to be freely varied). The best-fit output spectra were 
searched in the space of $g_{a\gamma}$ versus $m_a$.

On the basis of previous different searches for ALP signals and their 
constraints on $m_a$ and $g_{a\gamma}$ 
(e.g., ref. \cite{cast17,m87,aje+16,dmc20}),  
we considered a region of $10^{-10}\leq m_a\leq 10^{-6}$\,eV and 
$10^{-13}\leq g_{a\gamma}\leq 2\times 10^{-10}$\,GeV$^{-1}$. A key factor 
limited the size of this region was the large amount of computing time 
required.  The resulting $\chi^2$ values for the searched $m_a$ and 
$g_{a\gamma}$
region are shown in the left panel of Figure~\ref{fig:fit}. The minimum
$\chi^2_{\rm min}$ value is $\simeq$19.6 (26 degrees of freedom), and 
the 2$\sigma$ confidence areas ($\chi^2-\chi^2_{\rm min}\leq 9.21$) 
are marked
with red-dashed curves, helping locate the areas that provide good fits. 
There are basically two areas: (1) the lower right 
part and (2) the small narrow part in the range of $10^{-10}<m_a<10^{-8}$\,eV 
in the upper left corner. However, by checking the model spectra in the areas,
we found that in area (1) the model spectra do not contain any line-like
structures. The `good' results in the area are because of good fitting to
the continuum part of the observed spectrum. We note that if we use a 
LogParabola function to fit the observed spectrum, the best-fit 
$\chi^2\simeq 32.6$. 
\begin{figure}[tbp]
\centering 
\includegraphics[width=.45\textwidth]{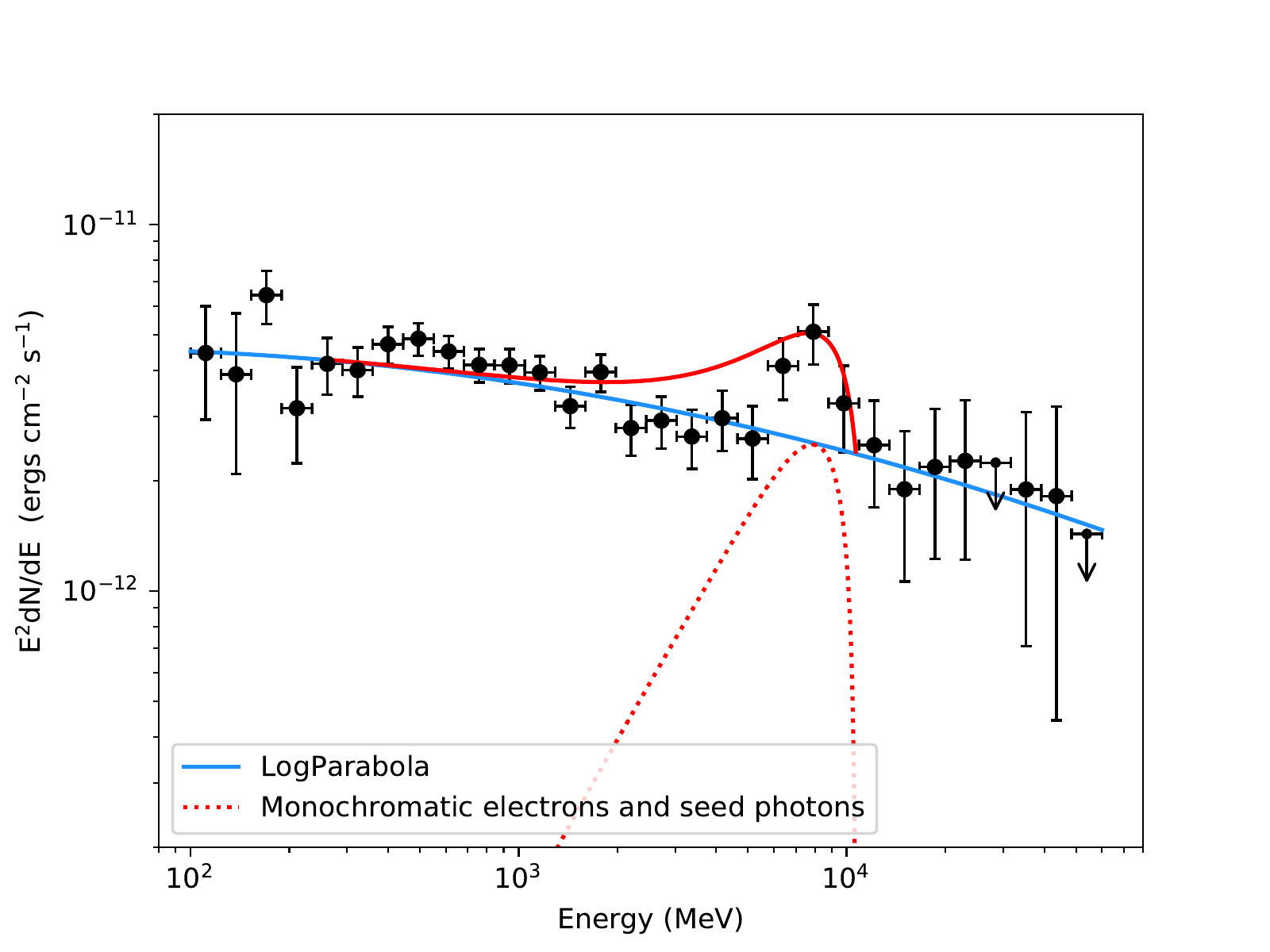}
\caption{A test fit to the line-like feature with the inverse Compton scattering
process from monochromatic electrons and seed photons.
\label{fig:lfit}}
\end{figure}

On the other side, model spectra in area (2) generally
contain a line-like structure. We thus re-did fitting for the area on much
fine grids. The results are shown in the middle panel of Figure~\ref{fig:fit}. 
The best-fit $\chi^2_{\rm min}\simeq 17.2$, and an example of the model 
spectrum is shown in the right panel of Figure~\ref{fig:fit}.  However
the good-fit areas are a few narrow lines along a given $g_{a\gamma}$ value.
In other words, for a given $m_a$ value, $g_{a\gamma}$ determines the position
and shape of the line-like structure around 7~GeV and thus whether a good 
fit can be provided to the observed spectrum. We show a set of model spectra
with $m_a\simeq 10^{-9}$\,eV and a set of step values of $g_{a\gamma}$ in our 
fitting in Figure~\ref{fig:fit}. As 
$g_{a\gamma}$ increases, the line-like structure shifts towards the low energy
and turns to be comparably broad and shallow, causing the fits to be worse.
Therefore even in area (2), there are only a few set of $g_{a\gamma}$
values can provide good fits to the observed spectrum.

As different combinations of the parameters can also provide fits to the SED of
B0516$-$621, $B_T^0=1$\,G is also an acceptable value (Figure~\ref{fig:bspec}).
For this $B_T^0$ value, $\delta=14.2$ and 
$R=9.5\times 10^{16}$~cm, and thus
$\Gamma_b\simeq 14.2$, $\theta\simeq 1/\Gamma_b\simeq 4.0^{\circ}$,
and $r_0\simeq 1.3\times10^{18}$\,cm. Setting these parameters, we re-did
the spectrum fitting calculations. The results are shown in 
Figure~\ref{fig:b1g}. We found that the $\chi^2$ distribution results
in the $g_{a\gamma}$ versus $m_a$ space have
a similar pattern
to those when $B_T^0=0.7$\,G, but are generally upshifted to 
larger $g_{a\gamma}$
values. In particular, the area that can provide acceptable fits is shifted
to $g_{a\gamma}$ values slightly greater than 10$^{-9}$\,GeV$^{-1}$.

\section{Discussion}
\label{sec:dis}

From analysis of 10-year \fermi\ LAT data, we have found a line-like feature in
the \gr\ spectrum of B0516$-$621. This feature has a significance of 
2--3$\sigma$, and likely appears in the source's emission all the time.
Previously, a $\sim$4$\sigma$ narrow spectral feature was seen in blazar
Markarian~501 at VHE $\sim$3\,TeV, but only appeared in one of short 
observations (1.5\,hrs) covering a flaring event \cite{m501}. Because this TeV
feature is relatively wide (FWHM$\sim$1.4~TeV) and at the high end of 
the observed spectrum, an
extra component arising from electrons of a narrow energy distribution can
provide the explanation (see ref.~\cite{m501} for details). For comparison,
the line-like feature of B0516$-$621 is narrower and appears in nearly
the middle of a continuum. As a result,
we found that even the most extreme case, electrons with a monochromatic 
energy distribution upper-scattering seed photons with a monochromatic 
distribution (i.e., the inverse Compton scattering process),
would only provide
a poor fit to the line-like feature (Figure~\ref{fig:lfit}; however, see 
discussion for the deep Klein-Nishina regime scenario; ref. \cite{akm12}). 
Therefore the
radiation processes generally considered in the frameworks for blazar 
high-energy emission can hardly explain the line-like feature.

\begin{figure}[tbp]
\centering 
\includegraphics[width=.45\textwidth]{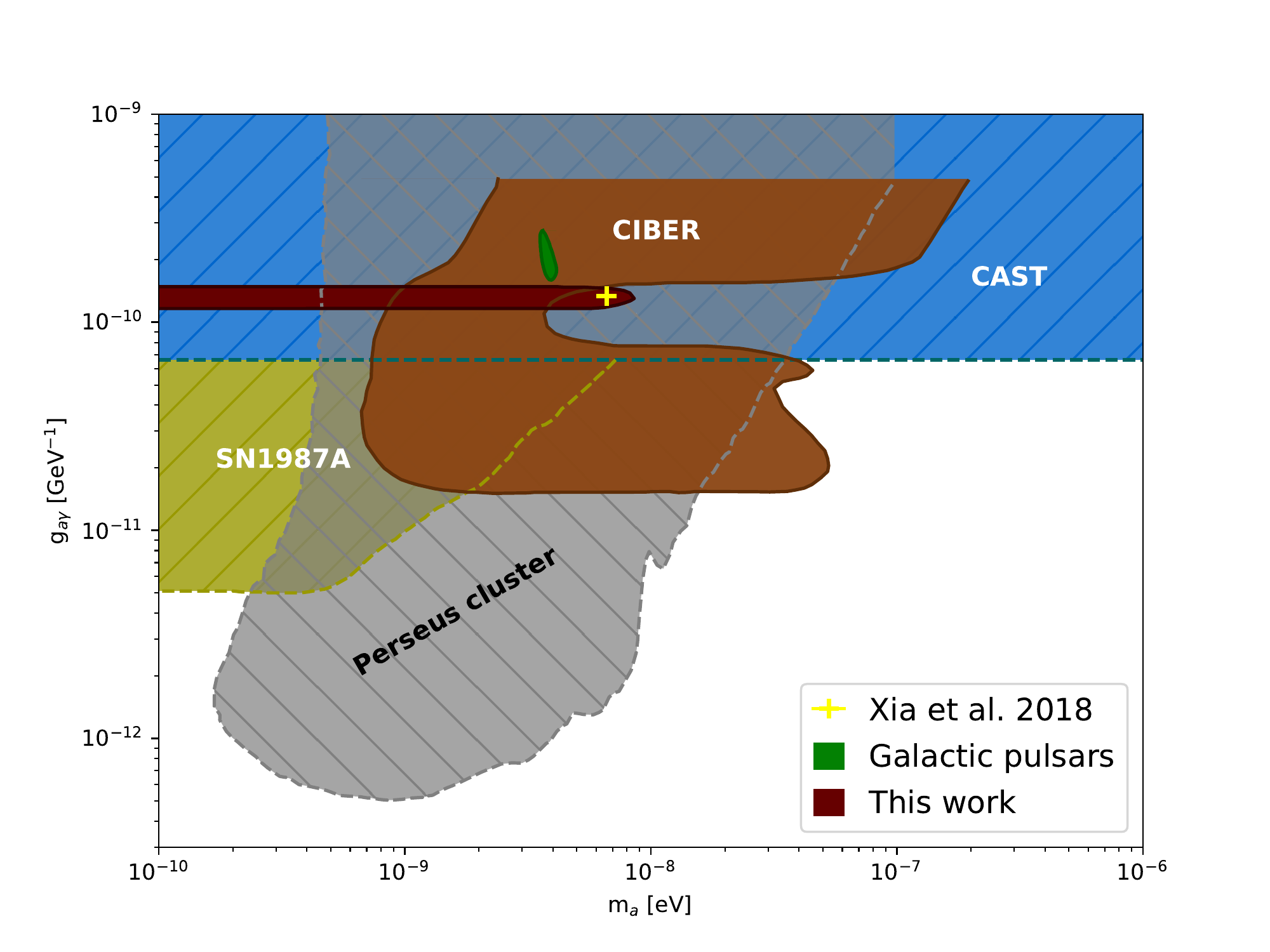}
\caption{Constraints on $g_{a\gamma}$ versus $m_a$ parameter space.
The shaded regions are excluded from several experiments or 
methods \cite{cast17,sn1987a,mal+18}, and the non-shaded regions 
(including a point marked by a plus and this work) are those preferred 
(see the text in Section~\ref{sec:dis} for details).
\label{fig:cons}}
\end{figure}

Instead, the line-like feature could be a signal due to photon-ALP oscillations.
We investigated this possibility by considering a jet case previously
in-detail discussed \cite{dmc20}. Large amount of the calculations were 
conducted to search in the $g_{a\gamma}$ versus $m_a$ 
parameter space that could give rise to a
line-like feature at $\sim$7~GeV. We found that in order to be able to
fit the line-like feature as well as
the continuum of the observed $\gamma$-ray spectrum,
a small region of $m_a\leq 10^{-8}$\,eV and 
$1.16\times 10^{-10}\leq g_{a\gamma}\leq 1.48\times 10^{-10}$\,GeV$^{-1}$ 
is preferred when $B_T^0=0.7$\,G. If considering a larger $B_T^0$ (=1\,G),
the region is shifted to even higher $g_{a\gamma}$ values, greater than 
$10^{-9}$\,GeV$^{-1}$. The regions are in tension with 
constraints (Figure~\ref{fig:cons}) obtained from other methods or 
experiments, in particular 
the CERN Axion Solar Telescope (CAST; ref.~\cite{cast17}).
However on the other hand, we note that the region we found with $B_T^0=0.7$\,G
partly overlaps with that derived from the CIBER experiment 
(Figure~\ref{fig:cons}; ref.~\cite{ciber}). Also, it is interesting to
note that our results are close
to a couple of other preferred
$m_a$ and $g_{a\gamma}$ ranges from studies of fine structures of
the \fermi\ LAT spectra, those of the bright supernova remnants \cite{xia+18}
and pulsars \cite{mch18}. The latter studies investigated photon-ALP 
oscillations in the Galactic magnetic field.

In any case, our studies of the \gr\ spectrum of the blazar B0516$-$621 present
another intriguing possibility for detecting ALP signals. It would be worth
finding similar features, thus allowing us to determine how common or rare 
such features would appear. Once other cases are found, the same studies
could be conducted, which helps identify reliable constraints on $m_a$ and
$g_{a\gamma}$. Based on our studies, we note that the spectral
resolution and sensitivity of LAT limit the searches for similar features;
sources need to be sufficiently bright (for which we may take B0516$-$621 as
a standard) in order to have a high-quality spectrum
showing identifiable features. In this respect, this B0516$-$621 case
points a direction of effort for future high-energy facilities that would have
high sensitivities and spectral resolutions for searching for similar features
and thus establishing more confident cases for detection of ALP signals.

\acknowledgments

We thank anonymous referee for helpful suggestions and M. Meyer for helping 
with the photon-ALP conversion code setting.
This research made use of the High Performance Computing Resource in the Core
Facility for Advanced Research Computing at Shanghai Astronomical Observatory.
This research was supported by the National Program on Key Research
and Development Project (Grant No. 2016YFA0400804) and the National Natural
Science Foundation of China (11633007, 11890692, U1931114, U1531130, U1831138).
Z.W. acknowledges the support by the Original
Innovation Program of the Chinese Academy of Sciences (E085021002).



\bibliographystyle{JHEP}

\providecommand{\href}[2]{#2}\begingroup\raggedright\endgroup
\end{document}